\documentclass{article}
\usepackage{arxiv}
\usepackage{graphics}
\usepackage{graphicx}
\usepackage{caption}
\usepackage{pgfplots}
\usepackage{subcaption}
\usepackage{xcolor}
\definecolor{arsenic}{rgb}{0.23, 0.27, 0.29}
\definecolor{zaffre}{rgb}{0.0, 0.08, 0.66}
\pgfplotsset{compat=1.18}
\AtBeginDocument{%
  }
\usepackage{listings}
\begin{document}
\title{persiansort : an alternative to mergesort inspired by persian rug
 }
 \author { parviz afereidoon\\
\texttt{afereidoon.p@gmail.com}
}
\setlength{\parindent}{3ex}
\begin {abstract}
    This paper introduces persiansort, new stable sorting algorithm  inspired by Persian rug. Persiansort does not have the weaknesses of mergesort under scenarios involving  nearly sorted and partially sorted data, also utilizing less auxiliary memory than mergesort and take advantage of runs. Initial experimental showed, this method is flexible, powerful and works better than mergesort in almost all types of data. Persiansort offers several advantages over merge methods make it a potential replacement.
\\
\\
\\
\end {abstract}

\maketitle

\section {Introduction}
 Stable sort algorithms sort equal elements in the same order that they appear in the input. This is important in applications where the original order of equal elements must be maintained. The following comparison-based sorting algorithms are stable by default: bubble sort, insertion sort and mergesort. \par
 Bubble sort is a simple and the slowest sorting algorithm which works on the principle of bubbling out the smallest or the largest element from the array depending on whether the array has to be sorted in descending or ascending order respectively \cite{Fe}.  Simple and easy to implement, time complexity of O(n) for already sorted dataset and use O(1) auxiliary space is some of benefit of bubble sort. But this algorithm works inefficiently for a large dataset.\par
 Insertion sort is a simple, in-place and an efficient sorting algorithm useful for small and nearly sorted lists. It inserts each element at it's appropriate position in the sorted sub-list and requires only a constant amount of additional memory space \cite{Rs}. This algorithm also works inefficiently for a large dataset.
 \par
 Mergesort is one of the most efficient and widely used sorting algorithms. Jon von Neumann proposed the mergesort algorithm in 1945 \cite{pk}. This algorithm is designed with the divide and conquer strategy. It has the best, worst and average running times as  O(nlogn) and uses additional memory as O(n). This algorithm is based on recursively dividing the input array into sub-arrays as often as needed until only one element remains in each sub-array and then combining the adjacent ordered sub-arrays from the bottom up to form one ordered sub-array. Mergesort is efficient for large datasets due to its O(nlogn) time complexity and is well-suited for sorting data stored on disk. Like any algorithm, mergesort has it's weaknesses:

\begin{itemize}
\item  Mergesort requires O(n) additional space for the temporary array used during merging.
\item  For small datasets, mergesort may be slower than simpler algorithms like insertion sort or bubble sort due to it's higher overhead.
\item Mergesort works inefficiently for nearly sorted data.
\item If the array is sorted, mergesort goes through the entire process...
\end{itemize} \par
 To solve some of the weaknesses of mergesort, natural merge sorts were first proposed by
Knuth \cite{pk}. Later, methods such as timsort by Peters and more recently  peeksort and powersort  by Munro and Wild were invented to increase the efficiency of mergesort in this field \cite{tp} \cite{Mw} \cite{Vj}\cite{Bn}\cite{an}. These methods are highly efficient for sorting data with runs (ascending and descending order). However, some weaknesses of mergesort still remain.
\\

 We seek a method that, while utilizing less auxiliary memory than mergesort and take advantage of runs, also does not have the weaknesses of mergesort under scenarios involving nearly sorted and partially sorted data.\par

\section {Page Setup}\par

Divide and conquer algorithms for sorting in the digital realm closely resembles design and weaving in the physical world. The algorithm's divisions resemble the warps, while the conquer stages correspond to the wefts interwoven with the warps of a textile.
In weaving, knots are occasionally employed alongside the warp and weft, as seen in rug weaving. In algorithms, knots are occasionally employed to enhance efficiency under varying conditions.
\begin{figure*}[h]
  \centering
  \includegraphics[width=15cm , height=5.0cm]{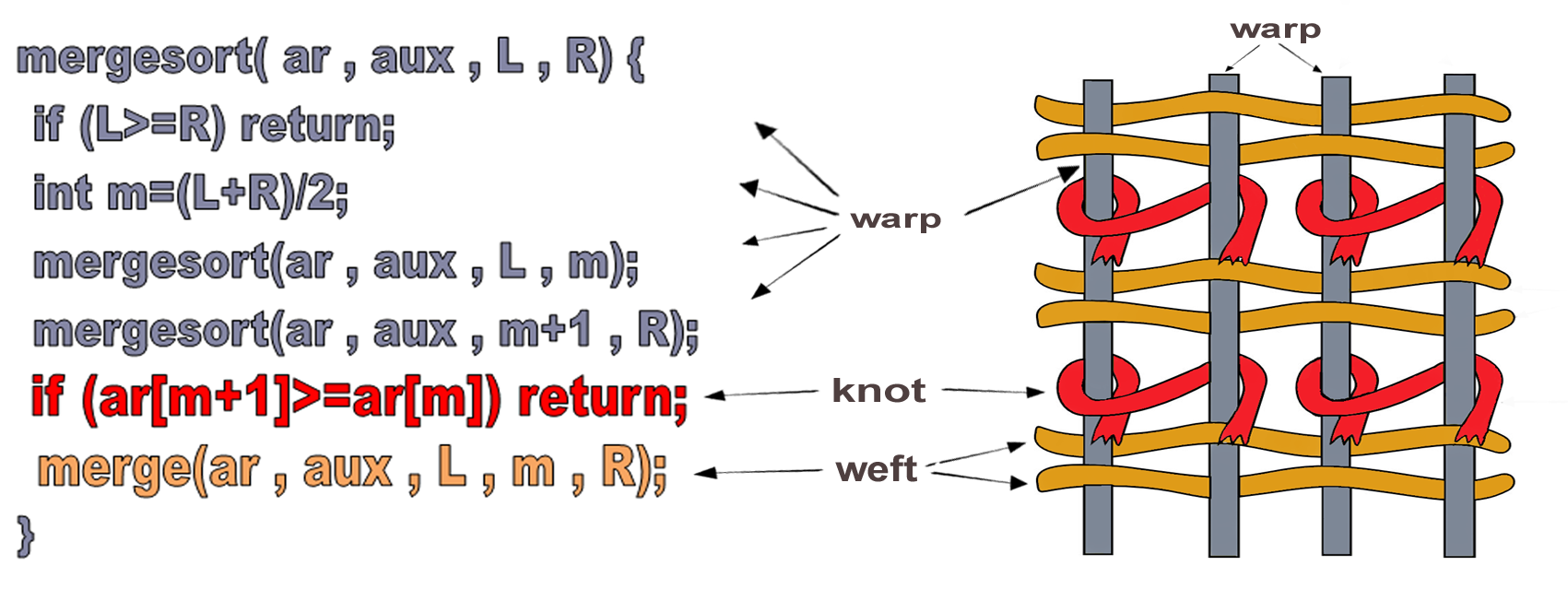}
  \vspace*{-2mm}
  \caption{{\color{arsenic}{\footnotesize divide and conquer algorithm and weaving.}}
}
  \label{fig:illu}
 \end{figure*}
A woven artifact such as a kilim can solely be constructed using warp and weft threads. Likewise, a divide and conquer algorithm can exist without knots. The variations in warp, weft, and knot types differ among rugs, akin to the distinctions in divisions, conquering stages, and knot kinds in sorting algorithms.\par
Consequently, we can draw inspiration from rug weaving to conceptualize a sorting algorithm. To this end, we reference Persian rug weavers who have not only perfected this art form and produce the finest handmade rugs globally, but also integrated rug weaving into their cultural identity.\par
The initial phase of rug weaving involves configuring the loom (warping), which varies according to the weaver's circumstances and the intended use of the rug. The warping process is executed in many methods and dimensions. The standard size is 12 square meters, but under specific circumstances, such as limited weaving time for nomads, a smaller size is used, akin to Gabbeh weaving\footnote{Gabbeh differs from rug. Gabbeh is smaller than rug and in contrast to rug, is woven without a pattern, relying on the weaver's mental design. It's knots are taller, and it's weft count is more, resulting in a softer texture.}. For expansive areas, the size may extend to hundreds of square meters\footnote{The Baharestan rug at Taq Kasra, originating from 600 AD and recently in 2007 largest rug is woven by about 1200 weavers from neishabour in 9 section, 5453 square meters, includes an estimated 2.2 billion knots.}. In ancient Persia, warping was utilized for the creation of double-sided rugs\footnote{Warping for the production of double-sided rugs has a lengthy historical background. Recent instances of it were discovered in the 16th century AD in Yazd by the Ghiyath al-Din carpet weaving workshops and currently, double-sided silk rugs are produced in the village of Doidokh.}. In the development of sorting algorithms, researchers employ various partitioning techniques, seeking to derive inspiration from this adaptability to create adaptable methods suitable for diverse situations and data kinds, while ensuring efficiency in sorting a broad spectrum of data.
The second point pertains to the rug weft, which, in this context, is utilized in varying sizes and materials based on the requirements of rug weaving. Additionally, we strive to ensure that the conquer phase in our algorithm achieves efficiency suitable for it's intended function.\par
The most crucial and labor-intensive stage in rug weaving is knotting\footnote{A 12 square meters rug contains over 5 million knots, requiring about 5500 hours (almost 3 years) for a weaver to complete. Exquisite rugs may possess up to double this quantity of knots.}. They employ many sorts of knots characterized by color, material, and knotting technique, demonstrating much sensitivity and delicacy in this process\footnote{The selection of knot materials and the longevity of natural dyes, comprising Approximately 200 fixed hues derived from madder, indigo, cochineal, saffron, etc, ensures that the colors of these rugs last for centuries. Prior to commencing the weaving process, skilled artisans occasionally chew a portion of the skein dyed yarn to verify its color durability..}. They also focus on the ratio of various elements in a woven item, such as color, material, design, and application, which they examine across a rug or carpet. In the realm of algorithms, adherence to this fraction correlates positively with increased algorithmic efficiency. Incorporating a strategy solely in one segment of a division for a certain condition may impair the algorithm's efficiency under other conditions. Employing an incongruous and uncoordinated technique relative to the division type can undermine the algorithm's worth and effectiveness.
We also include these factors in the design of our sorting algorithm, as the knot in the sorting algorithm is entirely affected by the partitioning method and the nature of the data being sorted. In the quicksort method, prior attempts to address the issue of duplicate data primarily concentrated on partitioning. However, by employing an appropriate and coordinated knot in a different position in eqsort, this problem has been more effectively resolved \cite{pa}.
\begin{figure*}[h]
  \centering
  \includegraphics[width=10cm , height=3.3cm]{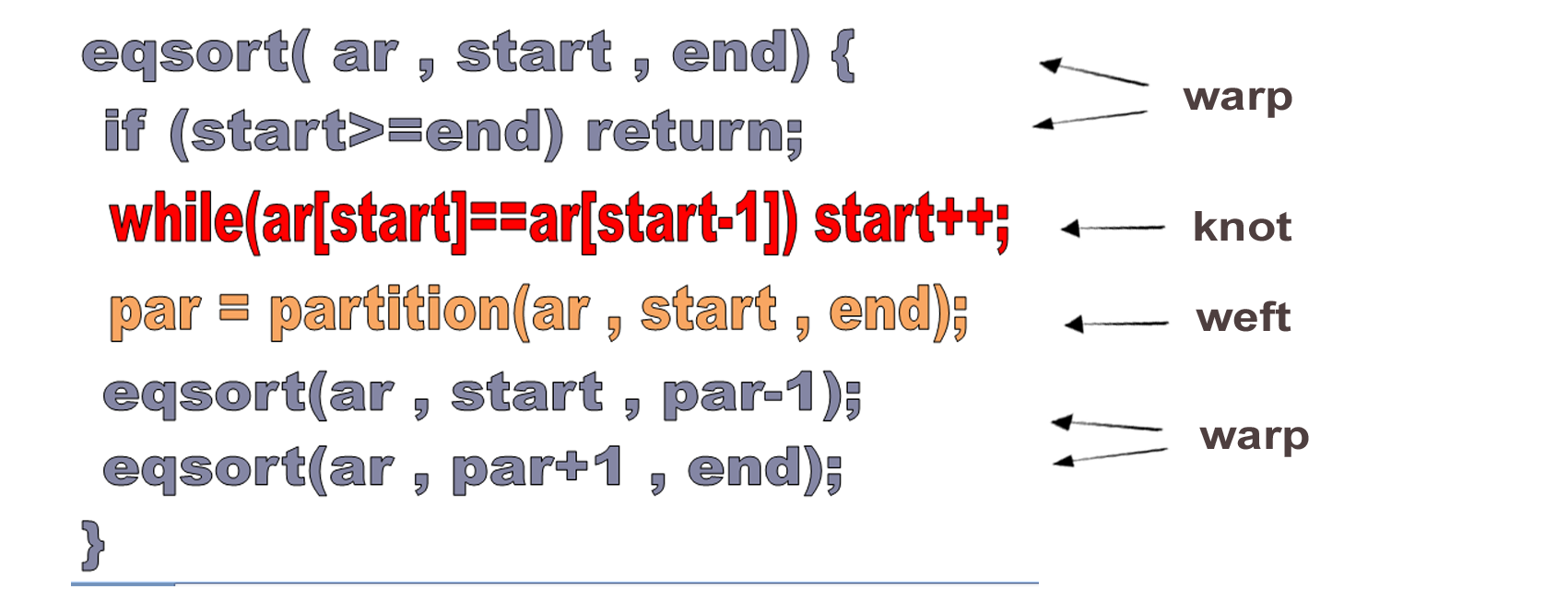}
  \vspace*{-2mm}
  \caption{\color{arsenic}{{\footnotesize eqsort algorithm and it's knot, weft and warps.\\}}
}
  \label{fig:illu}
 \end{figure*}
 \\
 This example illustrates that, akin to it's significance in rug weaving, the knot is likewise crucial and impactful in the design of a divide and conquer algorithm.
\section {New Algorithm Design}\par
The initial phase in developing the new algorithm is the division or warping stage. We select the top-down warp configuration for this purpose. In the top-down mergesort technique, the data is bifurcated at each stage but we do this with more flexibility so can make more effective knots.
To achieve this objective, we divide the data into further segments ("wp" segments). The number of segments (wp) can even change in different conditions and parts of the algorithm.
We obtain the size of each section from the following function: $((end-start+1)/wp)+1$ .\par
In persiansort, warping (divide step) can transition from wp=4 to wp=n+1 where "n" is the size of the original dataset. Alternative functions for warping may also be utilized.\par
You may alter the function at various stages of division.
Figure 3 illustrates the divide and conquer methodology.
In this instance, during the initial phase of division, a quantity of data is transferred to auxiliary memory (where "wp" is defined as wp=4). Subsequently, we have two options: we can either utilize the copied location of the original data as auxiliary memory or execute the remaining division steps directly on the auxiliary memory itself. We choose the second technique because it is more straightforward. At each phase of warping, the knot may take advantage of runs.

\begin{figure*}[h]
  \centering
  \includegraphics[width=17cm , height=3.3cm]{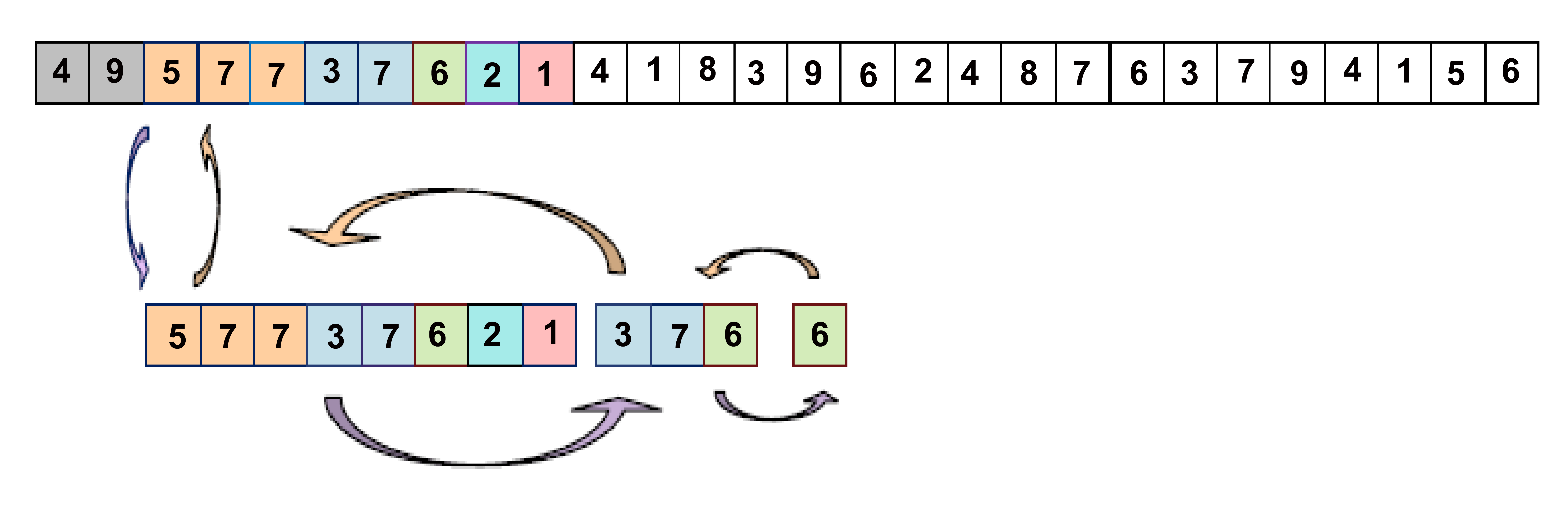}
  \vspace*{-2mm}
  \caption{\color{arsenic}{{\footnotesize warping or dividing stage}}
}
  \label{fig:illu}
 \end{figure*}
The second stage involves utilizing and identifying a weft suitable for this specific warp configuration or segmentation. To interlace the weft with the warps of this rug, a function is required to sort the final data after final division while concurrently uniting the various components throughout each conquest phase. We hereby define the jumpinsert function:
{\color{zaffre}
\begin{lstlisting}

  1:   void jumpinsert(int ar[],int start,int i,int aux[],int aux_start,int j){
  2:           int jump=j-aux_start+1; int temp;
  3:           while (j>=aux_start){
  4:              temp=aux[j];
  5:              while (i>=start && ar[i]>temp) {
  6:                 ar[i+jump]=ar[i];
  7:                 i--;
                  }
  8:              ar[i+jump]=temp;
  9:              jump--; j--;
               }
        }
 \end{lstlisting}\par
}
\begin{figure*}[h]
  \centering
  \includegraphics[width=15cm , height=11cm]{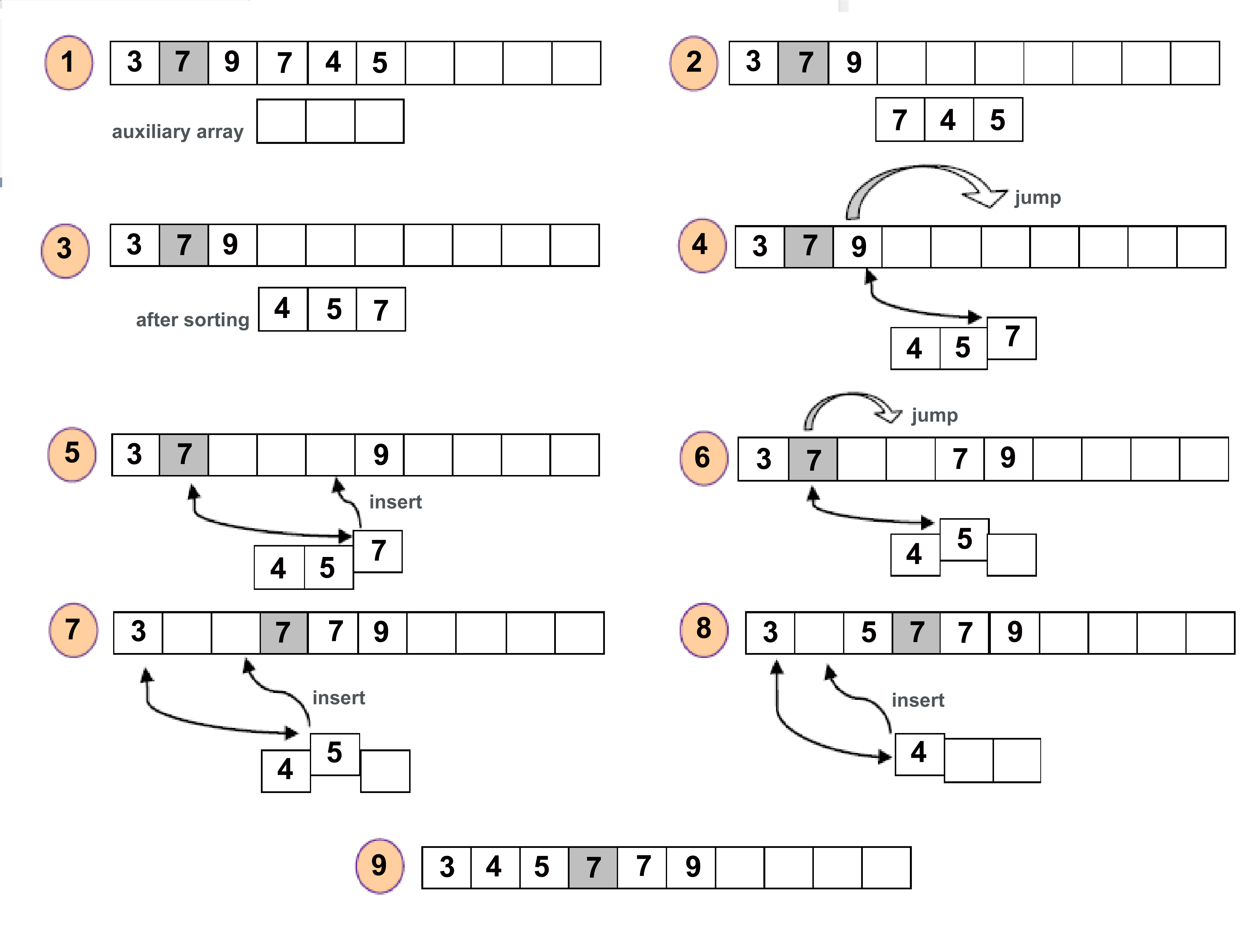}
  \vspace*{-2mm}
  \caption{\color{arsenic}{{\footnotesize jumpinsert function (weft)}}
}
  \label{fig:illu}
 \end{figure*}
Figure 4 illustrates a segment of the algorithm's operational methodology. The initial phase presents the overview. During the second phase, the data is transferred to auxiliary memory. Stage 3 is the phase in which sorting occurs. From Stage 4 forward, the operation of the jumpinsert function is illustrated, initiating the return of sorted data to the primary or preceding memory (the memory in each segment of the division is primary for the memory below it and auxiliary for the memory above it). Initially, the value of the largest auxiliary memory number is compared to that of the largest main memory number to return the data. If the data in main memory was larger, a jump is executed; otherwise, an insertion operation is performed. This graphic illustrates two data points with a value of 7, demonstrating the stability of the persiansort algorithm.\\
{\color{zaffre}
\begin{lstlisting}
10:  int auxiliary_size(int n, int wp){
11:      int m=n/wp;
12:      if(m==0) return 2;
13:       else return 1+m+auxiliary_size(m , wp);
      }

14:  void reverse (int ar[],int s,int e){
15:         while(s<e) {
16:              swap(ar[s],ar[e]);
17:              s++;e--;
            }
     }
18:   void warping(int ar[],int aux[],int start,int end,int aux_size,int wp){
19:          if(start>=end) return;
20:          int max_jump=((end-start+1)/wp)+1;
21:          int low,high,temp,j,i=start;
22:          while (ar[i+1]<ar[i] && i<end) i++;
23:          if(i>start) reverse(ar,start,i);
24:          while (i<end){
25:              while (ar[i+1]>=ar[i] && i<end) i++;
26:              if(i==end) return;
27:              if((i+max_jump)>end) max_jump=end-i;
28:              if((end-start+1)>aux_size)
                        {
29:                       for(j=1;j<max_jump+1;j++)
30:                             aux[j]=ar[i+j];
31:                       if(max_jump>1) warping(aux,aux,1,max_jump,aux_size,wp);
32:                       jumpinsert(ar,start,i,aux,1,max_jump);
33:                       i=i+max_jump;
                        }
34:               else
                        {
35:                        temp=end-i,low=end+1,high=end+max_jump;
36:                        for(j=low;j<high+1;j++)
37:                              ar[j]=ar[j-temp];
38:                        if(max_jump>1) warping(ar,ar,low,high,aux_size,wp);
39:                        jumpinsert(ar,start,i,ar,low,high);
40:                        i=i+max_jump;
                        }
             }
      }
41:  void persiansort(int ar[],int start,int end,int wp){
42:         int aux_size=auxiliary_size(end-start+1,wp);
43:         int aux[aux_size]{};
44:         if((end-start+1)<wp) aux_size=1;
45:         warping(ar,aux,start,end,aux_size,wp);
      }
\end{lstlisting}
}
The most crucial aspect is knotting many components of this algorithm, which can use of runs. To this end, we designate a pointer named i in line 21. Consequently, in each division, we retain the initial memory as pre-warp and by employing two knots in lines 22 and 25 the size of this segment (pre-warp) expands based on the data layout. The division process transpires subsequent to the operation of this knot, which can detect previously sorted data in various segments of the division. If the length of the sorted data exceeds wp, the second knot utilizes all of the runs. It also finds a substantial portion of the ordered data with a length inferior to wp.
Another benefit of these knots is that when all the data is pre-sorted, persiansort recognizes it with remarkable speed.
Additional knots may be incorporated into the persiansort algorithm as required, prior to line number 20, enabling hybridization with insertion sort when the data count falls below a specified threshold. Preliminary testing findings indicate that the application of this knot enhances speed in random data by around 10 percent. In addition to these, we can change of the max-jump during warping.\\
Line 27 governs the final stage of copying to the auxiliary array.\\
Line 28 determines whether we are transferring data from the main array to the auxiliary array or from the auxiliary array to itself.\\
Lines 42 and 43 quantify the auxiliary memory and allocate it accordingly.

\section {Time and Space complexity}\par

Due to the presence of the ascending and descending knots in lines 22 and 25 of the algorithm, the best-case time complexity of the algorithm is O(n), which occurs when the input data is already sorted.
Furthermore, when the data is nearly sorted, the algorithm maintains a linear time complexity of O(n).
Conversely, the worst-case time complexity arises when the data is completely random, such that the performance of knots is at its minimum. In this situation, the worst-case time complexity increases to O(nlogn).\\

The amount of memory required for this sorting method entirely depends on the selected value of wp and is determined by the function auxiliary-size. The required memory is given by the equation: $n/wp +n/(wp)^{2} +n/(wp)^{3}+...$  Nevertheless, if necessary, persiansort can be implemented in such a way that it requires only $ n/wp$ .\\

This algorithm demonstrates considerable flexibility and robustness, as optimal conditions for sorting various types of data can be achieved by adjusting a single parameter, denoted as wp.
From a theoretical standpoint, considering that for random datasets the number of comparisons in the persiansort algorithm can be approximately expressed by the relation $ ((wp)/2)n\log _{wp}n $ (assuming the impact of knots operations is neglected), the optimal value of wp is determined to be 4.
However, due to the influence of knots, the actual optimal value of wp is expected to differ.
Preliminary analyses and computations indicate that the optimal range for wp for the random dataset is estimated to be between 6 and 11(this value also depends on the parameter n).
Selecting any value within this range results in a performance variation of only about 0–5 percent.
Therefore, considering both performance stability and reduced memory consumption, the value 9 was chosen for wp.\\

For k-nearly sorted datasets, the parameter max-jump is typically set to a value approximately equal to k, and initial evaluations revealed that choosing an appropriate $ wp=n/(1.5\sqrt{k})$ value enhances the efficiency of the algorithm.

\section {Experimental}\par

To evaluate the effectiveness of persiansort and juxtapose this approach with other methods, we designated the wp value as 9. To evaluate the efficacy of this method and juxtapose it with alternative methods, we employ the same technique utilized for eqsort and sort a substantial quantity of arrays, analyzing the average time for varying data sizes \cite{pa}. For the purpose of comparing the sorting times of persiansort with other techniques, a large number of datasets  sorted. the data type has been set to double and the average sorting time is computed. The number of datasets is increased until a four-digit reproducibility is achieved, corresponding to an error of less than 0.1 percent.\\
All algorithms have been implemented in the C++ programming language and have been compiled with the GNU C++ Compiler 11.4.0 .  All tests have been performed on a PC with an i3-380M , 2 core , 2.53 GHz, with 4 GB memory and running Linux Ubuntu 22.04.\\

  The compared methods are selected as follows:\\
1.Persiansort method utilizing the original algorithm without hybridization, with wp=9.\\
    In persiansort for K-Nearly Sorted, we selected $ wp=n/(1.5\sqrt{k})$.\\
\\
2.	Mergesort2 (M2) with n/2 auxiliary memory:\\
\\
3.Mergesort1 (M1) with n auxiliary memory.\\
\\
4.	Timsort with minimum runs ranging from 32 to 64.\\
\\
5.	Insertion sort algorithm.\par
Sort times are displayed for different methods: T-M1 for mergesort1, T-M2 for mergesort2, T-TIMS for timsort, T-PER for persiansort, T-INS for insertion sort.

\section {Random Data}\par

The findings for random data indicate that the M1 algorithm is a little faster than other algorithm. M1 uses additional memory as O(n) and other results show that this method does not perform well in other conditions. Timsort is a hybrid, derived from merge sort and insertion sort. The novel persiansort approach can fully compete with Merge methods in random data, requiring significantly less auxiliary memory (n/wp).
  \begin{figure}[h]
   \begin{center}
   \begin{tikzpicture}[baseline]
   \begin{axis}[
   legend style={nodes={scale=0.65},at={(0.07,0.87)},anchor=west},
   legend cell align=left,
   xmax=10,xmin=1,
   width=9cm,height=6.5cm,
   ymin= 0.65,ymax=1.2,
   xticklabel style = {font=\small },
   yticklabel style = {font=\small },
   x label style ={at={(axis description cs:0.5,-0.2)}},
   xlabel=\emph{n (size of input)}, ylabel=\emph{Runtime ratio},
   xtick={1,2,3,4,5,6,7,8,9,10},
   ytick={0.7,0.8,...,1.2},
   xticklabels={$1000$,$2*10^{3}$,$5*10^{3}$,$10^{4}$,$2*10^{4}$,$5*10^{4}$,$10^{5}$,$2*10^{5}$,$5*10^{5}$,$10^{6}$},
   x tick label style={rotate=63 , anchor=east},
   ]
   \addplot [ black ,mark=square*]coordinates{(1,1) (2,1) (3,1) (4,1) (5,1)(6,1)(7,1) (8,1)(9,1)(10,1)};
   \addplot [black ,mark=triangle*] coordinates{(1,0.950) (2,0.940) (3,0.934) (4,0.933) (5,0.929)(6,0.926)(7,0.924)(8,0.920)(9,0.918)(10,0.916)};
   \addplot [red , mark=diamond*]coordinates{(1,0.925) (2,0.926) (3,0.950) (4,0.947) (5,0.949)(6,0.957)(7,0.944)(8,0.946)(9,0.965)(10,0.978)};
    \addplot [blue ,mark =*]coordinates{(1,0.919) (2,0.898) (3,0.924) (4,0.939) (5,0.924
   )(6,0.949)(7,0.966) (8,0.945)(9,0.967)(10,0.972)};
   \legend{\emph{T-M2 / T-M2},\emph{T-M1 / T-M2},\emph{T-TIMS / T-M2},\emph{T-PER / T-M2}}
   \end{axis}
   \end{tikzpicture}
   \vspace*{-2mm}
   \caption{\color{arsenic} Runtime ratio for mergesort1 , mergesort2 , timsort and persiansort methods in random data.}
   \label{fig:cs}
   \end{center}
   \end{figure}
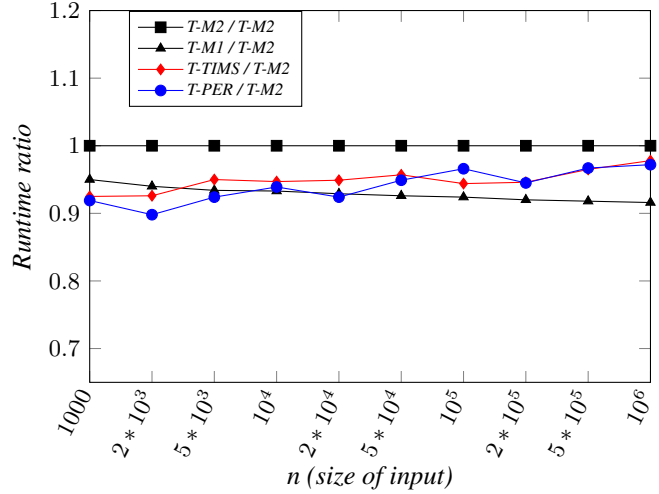
\section {Data with runs}
To evaluate the efficacy of the novel algorithm against several methods, we generate datasets including one million entries, ensuring that 60 percent of this data contains runs of size R (in ascending or descending order). For each measurement, we organize 1000 distinct sets under the stated parameters for each approach and calculate the average sorting time for each technique.
Figure 6  and other result illustrates that the effectiveness of the M1 technique declines significantly when the sort percentage of the dataset or the size of the runs increases, rendering this method an unsuitable option for this type of data. The remaining three approaches exhibit comparable speeds and efficiency. The persiansort approach has somewhat better performance as the sorting percentage and the size of the runs rise.

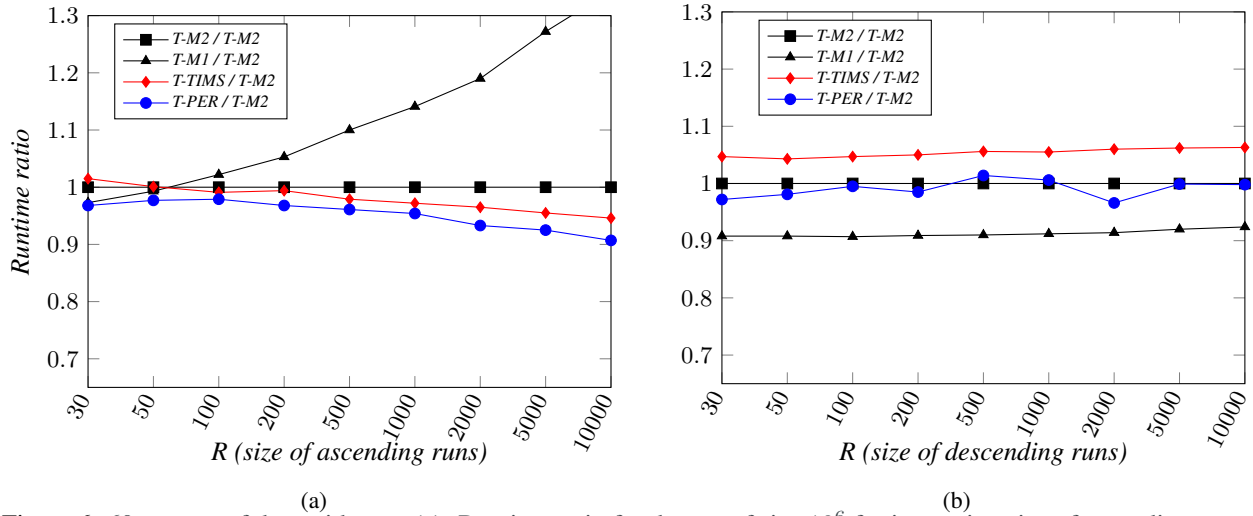
\begin{figure*}[h]
\begin{subfigure}[b]{0.5\linewidth}
\centering
 \begin{tikzpicture}
   \begin{axis}[
   legend style={nodes={scale=0.65},at={(0.05,0.85)},anchor=west},
   legend cell align=left,
   xmax=9,xmin=1,
   width=8.5cm,height=6.5cm,
   ymin= 0.65,ymax=1.3,
   xticklabel style = {font=\small },
   yticklabel style = {font=\small },
   x label style ={at={(axis description cs:0.5,-0.12)}},
   xlabel=\emph{R (size of ascending runs)}, ylabel=\emph{Runtime ratio},
   xtick={1,2,3,4,5,6,7,8,9},
   ytick={0.7,0.8,...,1.3},
   xticklabels={$30$,$50$,$100$,$200$,$500$,$1000$,$2000$,$5000$,$10000$},
   x tick label style={rotate=63 , anchor=east},
   ]
   \addplot [ black ,mark=square*]coordinates{(1,1) (2,1) (3,1) (4,1) (5,1)(6,1)(7,1) (8,1)(9,1)};
   \addplot [black ,mark=triangle*]coordinates{(1,0.973) (2,0.993) (3,1.022) (4,1.053) (5,1.100
   )(6,1.141)(7,1.190) (8,1.272)(9,1.343)};
   \addplot [red , mark=diamond*] coordinates{(1,1.015) (2,1.001) (3,0.991) (4,0.994) (5,0.979)(6,0.972)(7,0.965)(8,0.955)(9,0.946)};
   \addplot [blue ,mark =*]coordinates{(1,0.968) (2,0.977) (3,0.979) (4,0.968) (5,0.961)(6,0.954)(7,0.933)(8,0.925)(9,0.907)};
   \legend{\emph{T-M2 / T-M2},\emph{T-M1 / T-M2},\emph{T-TIMS / T-M2},\emph{T-PER / T-M2}}
   \end{axis}
   \end{tikzpicture}
   \caption{}
   \end{subfigure}%
   \begin{subfigure}[b]{0.53\linewidth}
   \centering
   \begin{tikzpicture}
   \begin{axis}[
   legend style={nodes={scale=0.65},at={(0.07,0.85)},anchor=west},
   legend cell align=left,
   xmax=9,xmin=1,
   width=8.5cm,height=6.5cm,
   ymin= 0.65,ymax=1.3,
   xticklabel style = {font=\small },
   yticklabel style = {font=\small },
   x label style ={at={(axis description cs:0.5,-0.13)}},
   xlabel=\emph{R (size of descending runs)}, ylabel=\emph{},
   xtick={1,2,3,4,5,6,7,8,9},
   ytick={0.7,0.8,...,1.3},
   xticklabels={$30$,$50$,$100$,$200$,$500$,$1000$,$2000$,$5000$,$10000$},
   x tick label style={rotate=63 , anchor=east},
   ]
   \addplot [ black ,mark=square*]coordinates{(1,1) (2,1) (3,1) (4,1) (5,1)(6,1)(7,1) (8,1)(9,1)};
   \addplot [black ,mark=triangle*]coordinates{(1,0.908) (2,0.908) (3,0.907) (4,0.909) (5,0.910
   )(6,0.912)(7,0.914) (8,0.920)(9,0.924)};
   \addplot [red , mark=diamond*] coordinates{(1,1.047) (2,1.043) (3,1.047) (4,1.050) (5,1.056)(6,1.055)(7,1.060)(8,1.062)(9,1.063)};
   \addplot [blue ,mark =*]coordinates{(1,0.972) (2,0.981) (3,0.995) (4,0.985) (5,1.014)(6,1.006)(7,0.966)(8,0.999)(9,0.998)};
   \legend{\emph{T-M2 / T-M2},\emph{T-M1 / T-M2},\emph{T-TIMS / T-M2},\emph{T-PER / T-M2}}
   \end{axis}
   \end{tikzpicture}
   \caption{}
   \end{subfigure}
   \vspace*{-7mm}
   \caption{\color{arsenic}60 percent of data with runs:(a)- Runtime ratio for  dataset of size $10^{6}$ for increasing size of ascending runs.
   (b)- Runtime ratio for increasing size of descending runs.}
   \label{fig:res6}
   \end{figure*}

\section {Nearly Sorted data}
We assess nearly sorted data in two sections to analyze and compare the persiansort approach with other techniques. In K-Nearly, each element is at most k positions away from its correct sorted position. In this kind of data, for persiansort, we designate the wp value as $ n/(1.5\sqrt{k})$.
In addition to the methodologies from prior stages, we also examine the insertion sort technique for K-Nearly.\\

 \begin{figure}[h!]
\centering
\begin{minipage}[b]{0.40\textwidth}
    \centering
    \begin{tabular}{ccccccc}
    \\ \cline{1-6}\\[0.1mm]
         K  & T-M1 & T-M2 & T-TIMS & T-INS & T-PER\\[0.5mm]\cline{1-6}\\[0.05mm]
         2     & 0.140 & 0.103 & 0.075  & 0.014 & 0.014 & \\
         5     & 0.163 & 0.151 & 0.109  & 0.019 & 0.018\\
         10     & 0.170 & 0.159 & 0.120  & 0.031 & 0.031&\\
         20     & 0.168 & 0.161 & 0.135  & 0.054 & 0.047&\\
         35     & 0.173 & 0.166 & 0.143  & 0.080 & 0.046&\\
         50     & 0.174 & 0.166 & 0.144  & 0.109 & 0.062&\\
         75     & 0.173 & 0.167 & 0.146  & 0.153 & 0.068&\\
         100     & 0.173 & 0.169 & 0.146  & 0.196 & 0.073&\\
         200     & 0.173 & 0.170 & 0.148  & 0.370 & 0.089&\\\cline{1-6}
              &  &  &   &  &&\\
              &  &  &   &  & &\\
              &  &  &   &  & &\\
    \end{tabular}
    \captionof{table}{\color{arsenic}Runtime for increasing k for k-nearly sorted data.}
    \label{tab:mytable}
\end{minipage}
\begin{minipage}[b]{0.45\textwidth}
    \centering
    \begin{subfigure}[b]{0.75\linewidth}
\centering
   \begin{tikzpicture}
   \begin{axis}[
   legend style={nodes={scale=0.65},at={(0.30,0.84)},anchor=west},
   legend cell align=left,
   xmax=9,xmin=1,
   width=8.0cm,height=6.7cm,
   ymin= 0,ymax=1.6,
   xticklabel style = {font=\small },
   yticklabel style = {font=\small },
   x label style ={at={(axis description cs:0.5,-0.1)}},
   xlabel=\emph{  k }, ylabel=\emph{Ratio},
   xtick={1,2,3,4,5,6,7,8,9},
   ytick={0.2,0.4,...,1.6},
   xticklabels={2,5,10,20,35,50,75,100,200},
   x tick label style={rotate=63 , anchor=east},
   ]
   \addplot[black ,mark=square*] coordinates{(1,1) (2,1) (3,1) (4,1) (5,1)(6,1)(7,1) (8,1)(9,1)};
   \addplot [ black ,mark=triangle*] coordinates{(1,1.354) (2,1.080) (3,1.070) (4,1.038) (5,1.040)(6,1.049)(7,1.036)(8,1.029)(9,1.021)};
   \addplot[red , mark=diamond*] coordinates{(1,0.728) (2,0.725) (3,0.756) (4,0.833) (5,0.859)(6,0.867)(7,0.872)(8,0.869)(9,0.874)};
   \addplot [ dashed , black,mark=*] coordinates{(1,0.133) (2,0.124) (3,0.192) (4,0.336) (5,0.484)(6,0.653)(7,0.916)(8,1.162)(9,2.182)};
   \addplot[blue ,mark =*] coordinates{(1,0.133) (2,0.124) (3,0.192)(4,0.294) (5,0.276)(6,0.375)(7,0.408)(8,0.435)(9,0.522)};
   \legend{\emph{T-M2 / T-M2},\emph{T-M1 / T-M2},\emph{T-TIMS / T-M2},\emph{T-INS / T-M2},\emph{T-PER / T-M2}}
   \end{axis}
   \end{tikzpicture}
   \end{subfigure}
    \caption{\color{arsenic}Runtime ratio}
    \label{fig:myplot}
\end{minipage}

\end{figure}

 The results in Figure 7 indicate that the persiansort approach significantly outperforms merge's methods.
An analysis of the two rapid methodologies reveals that:
\begin{itemize}
\item  The persiansort algorithm, akin to insertion sort, uses minimal auxiliary memory. persiansort utilizes memory equivalent to $1.5\sqrt{k}+3$.
\item The persiansort approach outperforms insertion sort for higher values of K, demonstrating significantly superior speed in certain instances, indicating that persiansort is among the most effective algorithms for sorting K-Nearly Sorted data.
\end{itemize}
A distinct kind of nearly sorted data is that in which the initial segment is ordered, exemplified by the addition of new data to an already sorted dataset, like online algorithms. We evaluated the novel persiansort approach against merging methods across various percentages of pre-sorted data.
\begin{figure}[h!]
\centering
\begin{minipage}[b]{0.40\textwidth}
    \centering
    \begin{tabular}{ccccc}
    \\ \cline{1-5}\\[0.1mm]
         $\%$pre-sorted  & T-M1 & T-M2 & T-TIMS  & T-PER\\[0.5mm]\cline{1-5}\\[0.05mm]
         65     & 0.168 & 0.122 & 0.113  & 0.118\\
         70     & 0.160 & 0.109 & 0.099  & 0.100\\
         75     & 0.151 & 0.093 & 0.087  & 0.087\\
         80     & 0.143 & 0.081 & 0.072  & 0.069\\
         85     & 0.135 & 0.069 & 0.060  & 0.055\\
         90     & 0.127 & 0.056 & 0.046  & 0.037\\
         93     & 0.123 & 0.049 & 0.039  & 0.029\\
         96     & 0.118 & 0.041 & 0.031  & 0.021\\
         99     & 0.113 & 0.035 & 0.024  & 0.013\\\cline{1-5}
              &  &  &   &   \\
              &  &  &   &   \\
              &  &  &   &   \\
              &  &  &   &   \\
    \end{tabular}

    \captionof{table}{\color{arsenic}Runtime for dataset of size $10^{6}$.}
    \label{tab:mytable}
\end{minipage}
\begin{minipage}[b]{0.45\textwidth}
    \centering
   \begin{subfigure}[b]{0.75\linewidth}
\centering
 \begin{tikzpicture}
   \begin{axis}[
   legend style={nodes={scale=0.65},at={(0.90,0.86)},anchor=east},
   legend cell align=left,
   mark options={scale=0.8},
   xmax=9,xmin=1,
   width=8.3cm,height=6.8cm,
   ymin= 0.1,ymax=1.5,
   xticklabel style = {font=\small },
   yticklabel style = {font=\small },
   x label style ={at={(axis description cs:0.5,-0.1)}},
   xlabel=\emph{    pp (percent of pre-sorted)}, ylabel=\emph{Ratio},
   ylabel style = {font=\small },
   xtick={1,2,3,4,5,6,7,8,9},
   ytick={0.2,0.4,...,1.6},
   xticklabels={65,70,75,80,85,90,93,96,99},
   x tick label style={rotate=63 , anchor=east},
   ]
   \addplot [black ,mark=square*] coordinates{(1,1) (2,1) (3,1) (4,1) (5,1 )(6,1)(7,1 )(8,1 )(9,1 )};
   \addplot[black ,mark=triangle*] coordinates{(1,1.378) (2,1.464) (3,1.620) (4,1.762) (5,1.956)(6,2.288)(7,2.525)(8,2.863)(9,3.278)};
   \addplot[red , mark=diamond* ] coordinates{ (1,0.926) (2,0.913) (3,0.936) (4,0.890)(5,0.863)(6,0.829 )(7,0.797 )(8,0.755 )(9,0.687 )};
   \addplot[blue ,mark =*] coordinates{(1,0.974) (2,0.915) (3,0.930) (4,0.842) (5,0.797)(6,0.663)(7,0.588 )(8,0.505 )(9,0.370 )};
   \legend{\emph{T-M2 / T-M2},\emph{T-M1 / T-M2},\emph{T-TIMS / T-M2},\emph{T-PER / T-M2}}
   \end{axis}
   \end{tikzpicture}
   \end{subfigure}%
    \caption{\color{arsenic}Runtime ratio}
    \label{fig:myplot}
\end{minipage}
\hfill

\end{figure}

Figure 8 illustrates that as the amount of pre-sorted data increases, persiansort significantly outperforms other approaches, achieving nearly double the speed of mergesort methods at large pre-sorted percentages.

\section {Conclusion}
The novel persiansort approach, inspired by Persian rug, demonstrates superior efficiency compared to previous methods in numerous instances. We utilized the original framework of persiansort to evaluate this method against alternative approaches, while it's inherent flexibility allows for enhancements in each domain. One of the key advantages of persiansort is it's ability to adapt to different types of data and makes it a versatile tool that can be applied to many different scenarios involving datasets rich in runs and nearly sorted data. It outperforms other stable methods and appears to be an appropriate alternative to mergesort in numerous instances.\\
\bibliographystyle{plain}
\bibliography{pe}
\appendix

 -------------------------------------------------------------------------------------------------------------------\\

 Mergesort1 (M1) algorithm.\\
{\color{arsenic}
\begin{lstlisting}
   mergesort(int a[],int b[],int start,int end){
      if(start>=end) return;
      if(descending==true) {reverse(a,start,end); return;}
      int m=(start+end)/2;
      mergesort(b,a,start,m)
      mergesort(b,a,m+1,end)

      if(b[m+1]>=b[m]) {coppybtoa(a,b,start,end); return;}
      merge(a,b,start,m,end);
   }
\end{lstlisting}
}
Mergesort2 (M2) algorithm \\
{\color{arsenic}
\begin{lstlisting}
   mergesort(int ar[],int aux[] ,int start,int end){
      if(start>=end) return;
      if(descending==true) {reverse(ar,start,end); return;}
      int m=(start+end)/2;
      mergesort(ar,aux,start,m)
      mergesort(ar,aux,m+1,end)

      if(ar[m+1]>=ar[m]) return;
      merge(ar,aux,start,m,end);
   }
\end{lstlisting}
}
\end{document}